
\voffset = -.2 in
\hoffset = -.2in
\headline={\ifnum\pageno=1\firstheadline\else
\ifodd\pageno\rightheadline \else\leftheadline\fi\fi}
\def\openone{\leavevmode\hbox{{\sevenrm 1}\kern -4.1pt 1}}
\def\firstheadline{\hfil}
\def\rightheadline{\hfil}
\def\leftheadline{\hfil}
        \footline={\ifnum\pageno=1\firstfootline\else\otherfootline\fi}
\def\firstfootline{\rm\hss\folio\hss}
\def\otherfootline{\hfil}

\font\tenrm=cmr10
\font\tenit=cmti10
\font\elevenbf=cmbx10 scaled\magstep 1
\font\elevenrm=cmr10 scaled\magstep 1
\font\elevenit=cmti10 scaled\magstep 1

\font\sevenrm=cmr7
\nopagenumbers
\line{\hfil }
\hsize=6.0truein
\vsize=8.5truein
\parindent=3pc
\baselineskip=10pt
\centerline{\elevenbf HANDLE OPERATORS IN R.C.F.T.}
\vglue 5pt
\vskip 24pt
\centerline{\tenrm M. CRESCIMANNO}
\baselineskip=13pt
\centerline{\tenit Center for Theoretical Physics, M.I.T., 77 Mass. Ave.}
\baselineskip=12pt
\centerline{\tenit Cambridge, MA. ~~02139-4307}

\vglue 0.8cm
\centerline{\tenrm ABSTRACT}
\vglue 0.2cm
{\rightskip=3pc
 \leftskip=3pc
 \tenrm\baselineskip=12pt
 \noindent For the series associated to a
group or coset R.C.F.T. there is a simple universal form
for the inverse of the handle operator in the ring of fusions.
These formulae may be easily understood from the quantization
of the associated Chern-Simons theory.}

\vglue 0.8cm
\line{\elevenbf 1. Introduction\hfil}
\bigskip
\baselineskip=14pt
\elevenrm
Among the data that defines a rational conformal
field theory (RCFT), the fusion algebra plays a central
role$^1$. It restricts the form of the modular representations
and fixes a convenient basis for discussing the dimension
of the space of blocks on higher genus ($g>0$). Let
roman letters label the integrable representations
of a group or coset RCFT and ${\cal O}_j$ be the
operators of fusion
$$ {\cal O}_i {\times} {\cal O}_j = {\rm N}_{ij}^k {\cal O}_k \ \ .
\eqno(1.1)$$
As described in Ref.[2,3], the dimension of the space of
conformal blocks in genus $g>0$ is given by
$$ {\rm dim}~{\cal H}^g = Tr(K^{g-1})\ ,  \eqno(1.2)$$
where $K$ is a matrix in the
space of integrable representations and is
a particular linear combination of the
operators ${\cal O}_j$
$$ K =  \sum_{l}^{}~Tr({\cal O}_l) ~{\cal O}_l . \eqno(1.3)$$
This is a very general characterization of $K$. For a gaussian
model (i.e. one composed of any number of non-interacting
massless bosons)
$K$ is the unit matrix scaled by the number of integrable representations
(a general property of $K^{-1}$ is $Tr(K^{-1}) = 1$.)
For the case of
group or coset RCFT, $K$ has no obvious classical group-theoretic
interpretation. That is, for example, given $G_k$ and forming
$K$ of Eq.(1.3) for this theory
it is clear that $K$ depends on the
level $k$ and on the group theory of $G$
in a complicated way.

The purpose of this note is to demonstrate that
the inverse matrix, $K^{-1}$ does admit a simple interpretation in terms
of classical group-theoretic ideas. Indeed, we show that
$K^{-1}$ for a given group or coset RCFT is
a ratio
$$ K^{-1} = w^2/{vol(k)}\ , 	\eqno(1.4)$$
where $w^2$ is a fixed linear combination of representations (depending
only on the group theory of $G$) and $vol(k)$ is the naive
volume of the moduli space of flat $G$-connections over the
torus and is a simple combinatorial factor that depends on the level $k$.
For example, for $SU(2)_k$ one finds
$$ K^{-1} = {{1}\over{2(k+2)}} \big [ 3{\cal O}_1 -{\cal O}_3\big]
\eqno(1.5)$$
for all level $k$. Here the subscripts refer to the
dimensions of the representations.

\vglue 0.6cm
\line{\elevenbf 2. Chern-Simons and the Handle Operator\hfil}
\vglue 0.4cm

Perhaps the simplest and most revealing way of understanding
formulae of this type is via the quantization of
Chern-Simons (CS) theory$^4$. Here, we sketch derivation of Eq.(1.4).
For more details see Ref.[5,6,7]. Recall that
the Hilbert space of CS theory is naturally isomorphic to
the space of conformal blocks of a CFT, and that one
convenient way to construct the Hilbert space of CS theory
is by quantizing the space, $\cal M$
of flat gauge connections over a given
surface. The moduli space $\cal M$ is not quite a manifold;
in general it contains a set of measure zero where
there are
cusps, self-intersections, disconnected points, and other
pathologies. Fortunately, we will only concern
ourselves here with the moduli space
of flat gauge connections over the torus
and this, for the classical
Lie groups, has only cusp-type singularities. Thus, for
that case it is enough to study the quantization of the
covering space $\hat{\cal M}$ of the moduli space
${\cal M} = {\hat{\cal M}}/W$, where the group $W$ is essentially
a realization of the Weyl group. As usual, the quantization on
$\hat{\cal M}$ proceeds via the symplectic form $\Omega = (k+c)Tr$
where $c$ is the quadratic casimir of the adjoint representation
and $Tr$ is the matrix through which the inner products
in the Cartan subalgebra are taken
(for the simply laced Lie algebras it is just the Cartan matrix; for the
non-simply laced case it is an appropriate symmetrization of the
Cartan matrix.) Call $\hat{\cal H}$
the quantum Hilbert space that results from the quantization
of covering space $\hat{\cal M}$. The Hilbert space
$\hat{\cal H}$ is
isomorphic to the space of conformal blocks of
a gaussian model , and admits an action
of the covering group $W$. Call that group action $W_{\cal H}$.
As discussed in the literature, the Hilbert space $\cal H$ of the
CS theory is found by modding $\hat{\cal H}$ by this group action
\halign{\hskip 2.5in #&#&#&#&#\cr
&&\hskip .25in$\Omega$\hskip .25in&&\cr
&~${\hat M}$~&\rightarrowfill&~${\hat{\cal H}}$~\cr
\vbox to 24pt {\vskip 16pt\hbox{$W$}\vfil} & ~$\Biggr\downarrow$
&&~$\Biggr\downarrow$ & \vbox to 24pt {\vskip 16pt\hbox{$W_{\cal H}$}
\vfil}\cr
&&&&\cr
&~$M$~&&~${\cal H}$~&\cr}

The ''handle-squashing'' operator $K^{-1}$ is then just the
push-forward of the handle-squashing operator in the $\hat{\cal H}$
theory. Since the $\hat{\cal H}$ theory
is gaussian, we know that on
$\hat{\cal H}$ the $K^{-1}$ is the unit matrix divided by
${\rm dim}\hat{\cal H}$.
For example, in $G_k$ CS theory
the $K^{-1}$ of the Hilbert space associated to the
cover of moduli space over the torus is
$$ K^{-1}_{\hat{\cal H}} = \big{|}{{\Lambda_w}\over{(k+c)\Lambda_r}}
\big{|}^{-1}\  {\bf \openone}  \ , \eqno(2.1)$$
where $\Lambda_w$ is the co-root lattice and
$\Lambda_r$ is the root lattice of $G$.

Now, just as all the points
in $\cal M$ are invariant under the action of $W$, all the states
of $\cal H$ form a covariant multiplets under the
action of $W_{\cal H}$. As described in the literature,
gauge invariance of the entire partition function requires that
the the states in $\psi\in{\cal H}$ are alternating states,
that is, they satisfy
$w\psi = det(w)\psi~~~\forall w\in W_{\cal H}$.
There is a fundamental alternating operator $\Gamma$ which ''projects''
all the states in $\hat{\cal H}$ onto the states in ${\cal H}$
$$ \Gamma = {{1}\over{\sqrt{|W|}}} \sum_{w\in W}^{} det(w)\Pi_i
B_i^{(\omega_i,w\rho)} \eqno(2.2)$$
where $\rho = {{1}\over{2}}\Sigma_{\alpha>0} \alpha$ and
the inner product in the exponent of the raising operators $B_i$
are in terms of the co-root basis, $\{ \omega_i\}$. $|W|$ is the
order of the Weyl group. To compute modular
representations, fusions, etc. in the ring of operators on $\cal H$
one simply conjugates the corresponding quantity in the
$\hat{\cal H}$ by the operator $\Gamma$. Thus, by  pushing-forward
Eq.(2.1) we find
the handle-squashing operator for the case of $G_k$ to be
$K^{-1} ={{|W|}\over{{\rm dim}{\hat{\cal H}}}} \Gamma^{\dagger}
{\bf \openone}\Gamma$
which we may write (using the consequence of symmetry
$\Gamma^{\dagger}=-\Gamma$) as
$$ K^{-1}_{G} = - {{1}\over{vol(G_k)}} \Gamma^2  \ \ ,	\eqno(2.3)$$
where
$$ vol(G_k) = {{\big{|}{{\Lambda_w}\over{(k+c)\Lambda_r}}\big{|}}\over
{|W|}}   \eqno(2.4)$$
may be thought of as the naive symplectic volume of the
$\cal M$ and $\Gamma^2$ is, of course, invariant under the
action of the Weyl group and, for a given $G$, is a fixed linear combination
of representations and is thus independent of the level.

For cosets ''without fixed points'' there is also a simple formula
for the handle-squashing operator $K^{-1}$
of the form Eq.(2.3). For a simple coset $G_k/H_k$, where the
bonus currents act without fixed points
$$ K^{-1}_{G/H} = {|Z|}^2 (K^{-1}_G \otimes K^{-1}_H){\big{|}}_{G/H}
\eqno(2.5)$$
where $Z = Z_G\cap H$ is the common center and $|Z|$
is the number of elements it has. Since the center action
may be thought of as acting
via $Z\times Z$
as a further identification
in $\hat{{\cal M}}_G \times \hat{\cal M}_H$ and since it acts
freely, we see that collecting the factors we may represent
$K^{-1}_{G/H}$ again as a ratio of a particular group
theoretic part and a naive symplectic volume
of the coset's moduli space of flat connections
$vol(G_k/H_k) = vol(G_k)vol(H_k)/|Z|^2$.

\vglue 0.6cm
\line{\elevenbf 3. Examples and Summary\hfil}
\vglue 0.4cm
Using the above ideas we now list a few examples of these
explicit handle operator formulae
$$ K^{-1}_{U(1)} = {{1}\over{2k}} {\cal O}_1 $$
$$ K^{-1}_{SU(2)} = {{1}\over{2(k+2)}}(3{\cal O}_1-{\cal O}_3)$$
$$ K^{-1}_{SU(3)} = {{1}\over{3(k+3)^2}}(9{\cal O}_1-6{\cal O}_8
+3[{\cal O}_{10}+{\cal O}_{\bar{10}}]-{\cal O}_{27})$$
$$ K^{-1}_{SO(5)} = {{1}\over{4(k+3)^2}}(22{\cal O}_{(0,0)}-4{\cal O}_{(1,0)}
-7{\cal O}_{(0,2)}-2{\cal O}_{(2,0)}+6{\cal O}_{(1,2)}-3{\cal O}_{(3,0)}
-3{\cal O}_{(0,4)}+{\cal O}_{(2,2)})$$

{}~~

$$ K^{-1}_{SU(2)/U(1)} = {{1}\over{k(k+2)}}(3{\cal O}_1\otimes{\cal O}_1
-{\cal O}_3\otimes{\cal O}_1)$$
Note that for low level $k$ the operators in the above
expressions are to be understood as being
related to the integral representations modulo the action of the
'Weyl reflection about the $k$-line'$^{5}$.

We have described in general for group and simple coset
type RCFT's how formulae for $K^{-1}$ of this type arise.
The formulae have a natural interpretation in CS theory as
a ratio of a group-theoretic part and a naive volume
of moduli space.

There is an intriguing mathematical
connection between these ideas in the RCFT
context and $N=2$ theories. Also it seems
possible to explore properties
of the space of blocks in higher genus with
some of these techniques. Finally, these formulae
may be useful for better understanding the
classical ($k\rightarrow\infty$) limit of CS theory.
%

\vglue 0.6cm
\line{\elevenbf 4. Acknowledgements \hfil}
\vglue 0.4cm

We wish to thank S. Axelrod, K. Bardakci,
S. A. Hotes, H. J. Schnitzer and I. M. Singer
for conversations, and the
organizers of this conference.
This work was supported in part by funds provided
by the U.S. Department of Energy (D.O.E.) under contract
\#DE-AC02-76ER03069, and by the Division of Applied Mathematics of the
U.S. Department of Energy under contract \#DE-FG02-88ER25066.
\vglue 0.6cm
\line{\elevenbf 5. References \hfil}
\medskip
\item{1.} E. Verlinde, {\elevenit Nucl. Phys.\/} {\bf B300} (1988) 360.
\item{2.} H. Verlinde and E. Verlinde, "Conformal Field Theory and
Geometric Quantizatio," published in {\bf Trieste Superstrings} (1989), 422.
\item{3.} R. Bott, {\elevenit Surveys in Diff. Geom.\/} {\bf 1} (1991) 1.
\item{4.} E. Witten, {\elevenit Commun. Math. Phys.\/} {\bf 121} (1989) 351.
\item{5.} M. Crescimanno and S. A. Hotes, {\elevenit Nucl. Phys.\/}
{\bf B372} (1992) 683.
\item{6.} M. Crescimanno, {\elevenit Nucl. Phys.\/} {\bf B393} (1993) 361.
\item{7.} M. Crescimanno, {\elevenit Mod. Phys. Lett.\/} {\bf A8} (1993)
1877.

\vfill
\eject
\bye